\pdfoutput=1

\documentclass{PoS}

\usepackage{graphicx}

\usepackage{ifthen}

\newcommand{\eg}{\textit{e.g.}}
\newcommand{\etal}{\textit{et~al.}}

\newcommand{\Refeqn}[2][eqn:]{Equation~(\ref{#1#2})}

\newcommand{\reffig}[2][fig:]{Figure~\ref{#1#2}}

\newcommand{\refsec}[2][sec:]{Section~\ref{#1#2}} 

\newcommand{\ifmulticol}[2]{%
  \ifthenelse{\lengthtest{1.9\columnwidth<\textwidth}}{#1}{#2}%
}

\newcommand{\scfigwidth}{%
  \ifdim1.9\columnwidth<\textwidth%
    1.00\columnwidth\else0.60\columnwidth\fi%
}

\newcommand{\widefigwidth}{%
  \ifdim1.9\columnwidth<\textwidth%
    1.00\columnwidth\else0.80\columnwidth\fi%
}

\newcommand{\insertfig}[1]{%
    \includegraphics[keepaspectratio,width=\scfigwidth,
                     height=0.35\textheight]{#1}
}

\newcommand{\insertwidefig}[1]{%
    \includegraphics[keepaspectratio,width=\widefigwidth,
                     height=0.35\textheight]{#1}
}

\newcommand{\insertdoublefig}[2]{%
    \includegraphics[keepaspectratio,width=0.49\textwidth,
                     height=0.35\textheight]{#1}
    \hspace{\stretch{1}}
    \includegraphics[keepaspectratio,width=0.49\textwidth,
                     height=0.35\textheight]{#2}
}

\newcommand{\gae}{%
  \ensuremath{\lower 2pt \hbox{%
    $\, \buildrel {\scriptstyle >}\over {\scriptstyle \sim}\,$}%
    }%
  }
\newcommand{\lae}{%
  \ensuremath{\lower 2pt \hbox{%
    $\, \buildrel {\scriptstyle <}\over {\scriptstyle \sim}\,$}%
    }%
  }

\newcommand{\Leff}{\ensuremath{\mathcal{L}_{\textrm{eff}}}}
\newcommand{\Enr}{\ensuremath{E_{nr}}}

\newcommand{\Snr}{\ensuremath{S_{nr}}}
\newcommand{\See}{\ensuremath{S_{ee}}}
\newcommand{\Sone}{\ensuremath{S1}}
\newcommand{\Stwo}{\ensuremath{S2}}
\newcommand{\Soneave}{\ensuremath{\langle\Sone\rangle}}

\newcommand{\etaSone}{\ensuremath{\eta_{S1}}}

\title{XENON10/100 dark matter constraints:
       examining the \Leff\ dependence}

\ShortTitle{XENON10/100 dark matter constraints:
            examining the \Leff\ dependence}

\author{\speaker{Christopher SAVAGE}%
        \thanks{Based on work done in collaboration with
                Katherine Freese, Graciela Gelmini, and Paolo Gondolo.
                }\\
        Stockholm University\\
        E-mail: \email{savage@fysik.su.se}
        }

\FullConference{Identification of Dark Matter 2010\\
                July 26 - 30 2010\\
                University of Montpellier 2, Montpellier, France}

\date{\today}

\abstract{
The determination of dark matter constraints from liquid xenon direct
detection experiments depends upon the amount of scintillation light
produced by nuclear recoils in the detector, a quantity that is
characterized by the scintillation efficiency factor \Leff.  We examine
how uncertainties in the measurements of \Leff\ and the extrapolated
behavior of \Leff\ at low recoil energies (where measurements do not
exist) affect the constraints from experiments such as XENON10 and
XENON100, particularly in the light WIMP regions of interest for the
DAMA and CoGeNT experimental results.
} 

\begin{document}


\section{\label{sec:Intro} Introduction}

The DAMA annual modulation \cite{Bernabei:2010mq,Belli:2010idm} and
the excess low-energy events in CoGeNT \cite{Aalseth:2010vx} can be
explained by recoils from a light Weakly Interacting Massive Particle
(WIMP); see \eg\ Ref.~\cite{Hooper:2010uy}.
The dark matter interpretation of these signals can potentially be
confirmed or refuted by experiments such as XENON10 \cite{Angle:2009xb}
and XENON100 \cite{Aprile:2010um,Schumann:2010idm}, but the ability
of these experiments to probe light WIMPs is dependent on their
sensitivity to low energy recoil events in their detectors.  The
calibration of these detectors' energy scales (and thus knowledge of
their sensitivity to low-energy events) is dependent upon the
scintillation efficiency factor \Leff.  We examine here how the XENON
constraints depend upon various models for the poorly known \Leff.

Some of the work presented here is described in Ref.~\cite{Savage:2010tg},
although the analysis of the XENON10 results has been extended down to
lower recoil energies with a more careful treatment \cite{Savage:2011}.

\section{\label{sec:Leff} The \Leff\ Scintillation Efficiency Factor}

\begin{figure*}
  \begin{center}
  \insertfig{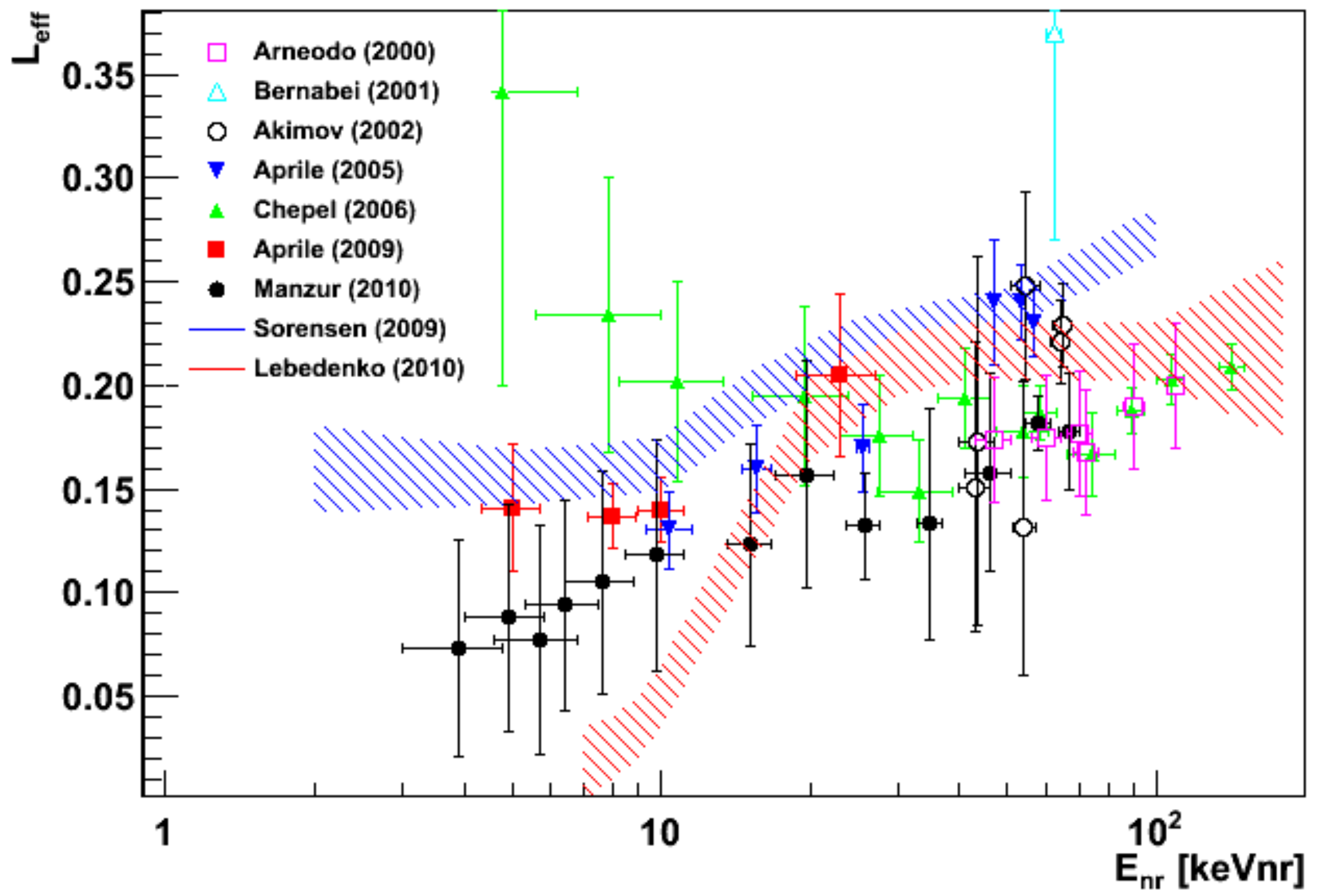}
  \end{center}
  \caption[\Leff\ measurements]{
    Various measurements of the scintillation efficiency factor \Leff.
    Figure courtesy of M.~Schumann.
    }
  \label{fig:LeffMeasurements}
\end{figure*}

\begin{figure*}
  \insertdoublefig{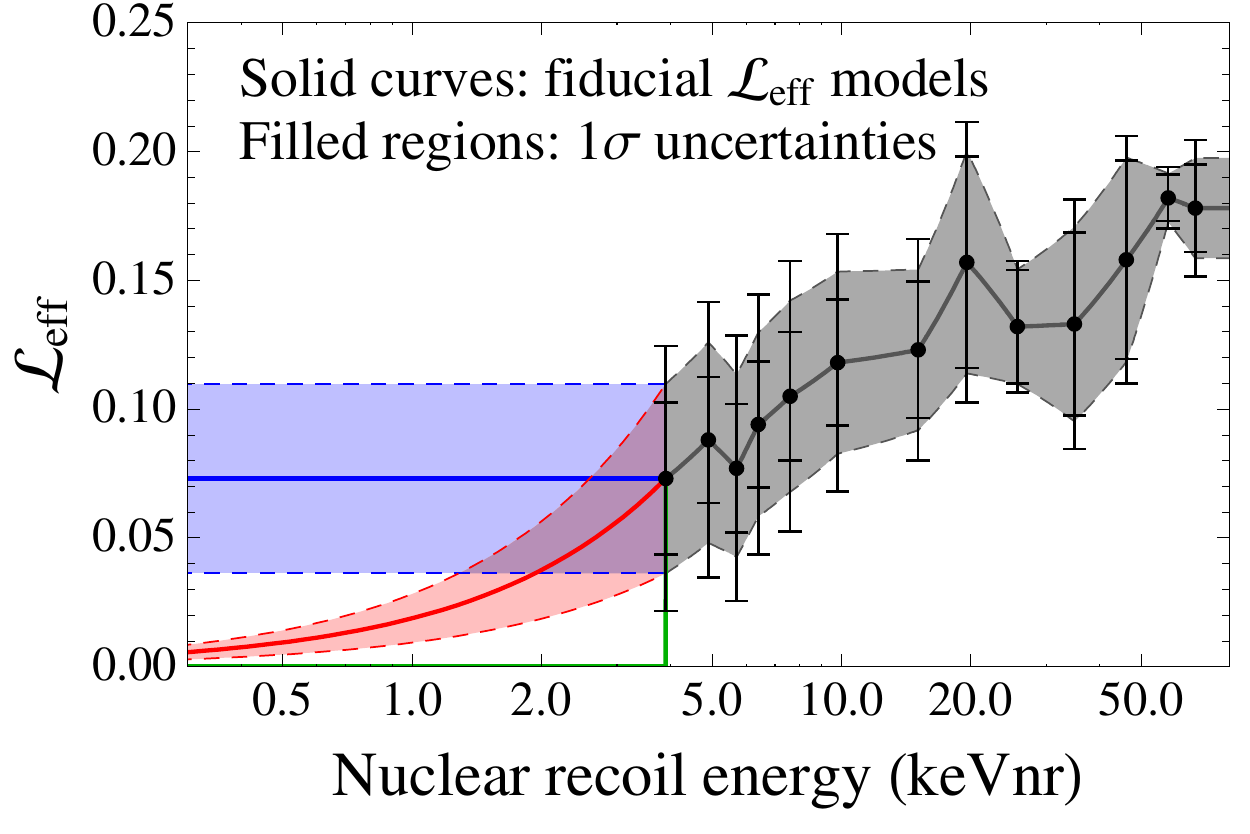}{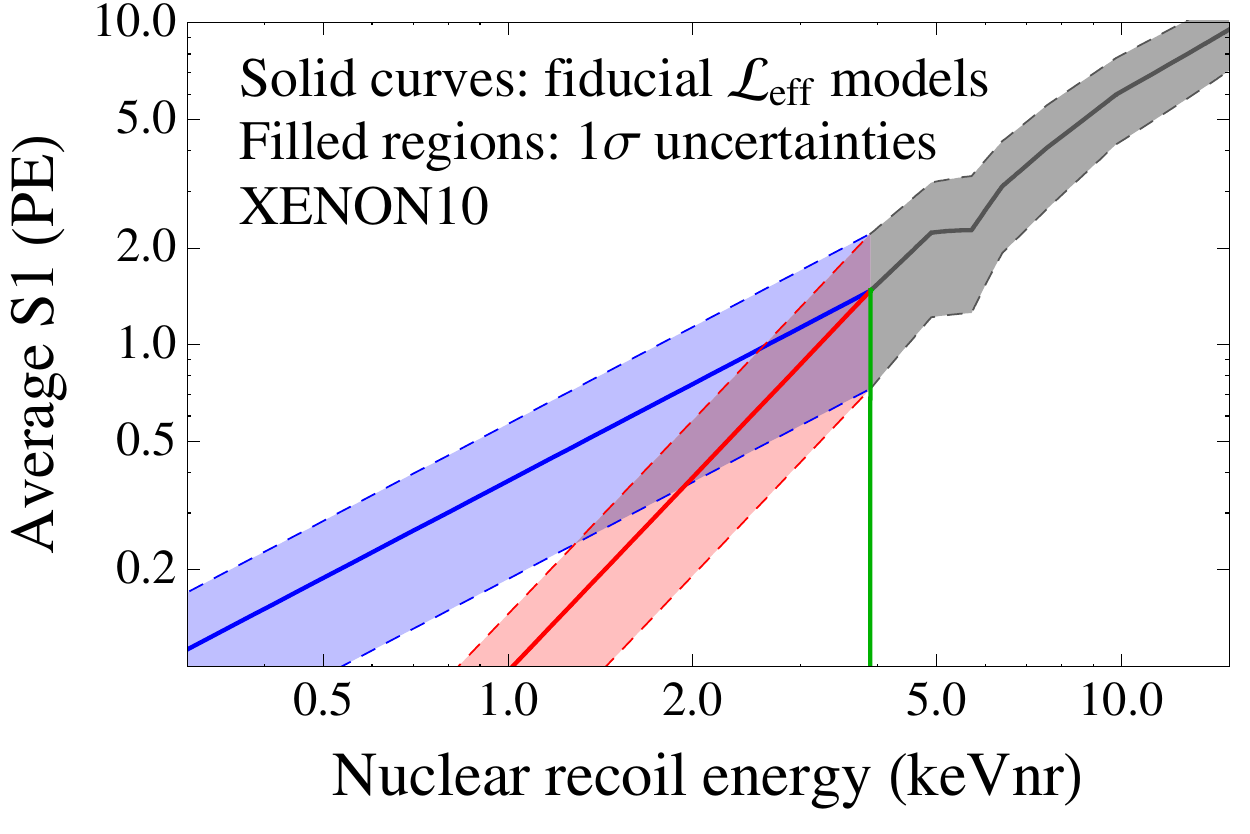}
  \caption[\Leff\ models]{
    (\textit{left})
    Assumed models for \Leff, based on the Manzur \etal\ measurements
    \cite{Manzur:2009hp} with three different extrapolations below
    \Enr~=~3.9~keV: constant (blue), linearly falling (red), and
    zero (green).  Solid curves indicate fiducial cases, while lighter
    regions indicate 1$\sigma$ uncertainties.
    (\textit{right})
    The expected average \Sone\ signals in XENON10 for these \Leff\ 
    models.
    XENON100 exhibits a similar (but not identical) behavior.
    }
  \label{fig:LeffModels}
\end{figure*}

Recoils in the XENON detectors produce two scintillation signals,
referred to as \Sone\ and \Stwo, resulting respectively from prompt
photons and ionization produced by the collision of a WIMP with a
xenon nucleus.
Interpretation of the XENON results requires the ability to reliably
reconstruct the nuclear recoil energy \Enr\ from the observed \Sone\ 
signal.
Calibration of the nuclear recoil energy dependence of \Sone\ 
often involves gauging the detector's response to electron recoils at
higher energies; parts of the detector's response (\eg\ the fraction of
scintillation photons that yield photoelectrons (PE) in the
photodetectors) are more easily determined in this case than with
nuclear recoils at lower energies.
Taking \Sone\ to be normalized to the number of PE,
\Sone\ and \Enr\ are related by an equation involving the higher
energy electron recoil calibrations:
\begin{equation} \label{eqn:S1}
  \Soneave = (\Snr / \See) \, \Leff(\Enr) \, L_y \, \Enr \, .
\end{equation}
Here, $L_y$ is the light yield in PE/keVee for 122~keV $\gamma$-rays.
\Leff(\Enr) is the scintillation efficiency of
nuclear recoils relative to 122~keV $\gamma$-rays in zero electric
field;  this factor is a function of the nuclear recoil energy.
Since there is an applied electric field in the experiment, which
reduces the scintillation yield by quickly removing charged particles
from the original interaction region,
two additional factors must be taken into account:
\See\ and \Snr\ are the suppression in the scintillation yield for
electronic and nuclear recoils, respectively, due to the presence of
the electric field in the detector volume.
The quantities \See, \Snr, and $L_y$ are detector dependent; \Leff\ is
not.
\Refeqn{S1} describes the \textit{average} \Sone\ signal (\Soneave);
the actual observed \Sone\ signals exhibits some random fluctuations
about this average as described in \refsec{Detection}.

A variety of \Leff\ measurements and estimates are shown in
\reffig{LeffMeasurements}.  These measurements are plagued by systematic
issues and give conflicting estimates of \Leff\ at low recoil energies;
see Ref.~\cite{Manalaysay:2010mb} for a discussion regarding the various
issues in these \Leff\ measurements.
There are two issues here:
(1) Which of the \Leff\ measurements should be used as a basis for
analyzing direct detection results? and (2) Measurements of \Leff\ 
have only been made at energies above some minimum; what is the
behavior of \Leff\ at low energies, where no measurements have as yet
been made?
We do not address the first issue (which set of measurements to use)
and simply choose the set of fixed-energy measurements that will give
the most conservative constraints: the Manzur \etal\ measurements
\cite{Manzur:2009hp,McKinsey:2010idm} (black points).
For the second issue (the behavior of \Leff\ at energies below the
existing measurements), we consider three extrapolations of \Leff\ 
below recoil energies of 3.9~keV (the lowest Manzur measurement):
(1) \Leff\ is constant,
(2) \Leff\ is linearly falling with recoil energy, and
(3) \Leff\ is strictly zero.
The XENON estimates of \Leff\ have been suggestive of the first
(constant) case, while the Manzur measurements are more suggestive of
the second (falling) case.  The third case leads to the most
conservative possible constraints by simply ignoring any contributions
from recoils with energies below 3.9~keV and is not necessarily a
realistic model of the low-energy \Leff\ behavior.
In addition to the dependence of the XENON constraints on these
low-energy extrapolations, we consider how the constraints vary within
the 1$\sigma$ uncertainties in the Manzur measurements.
Our \Leff\ models are shown in the left panel of \reffig{LeffModels},
while the right panel of the figure shows the corresponding average
\Sone\ signals in the XENON10 detector.

\section{\label{sec:Detection} Detection in XENON10/100}

\begin{figure*}
  \includegraphics[keepaspectratio,width=0.45\textwidth,
                   height=0.35\textheight]{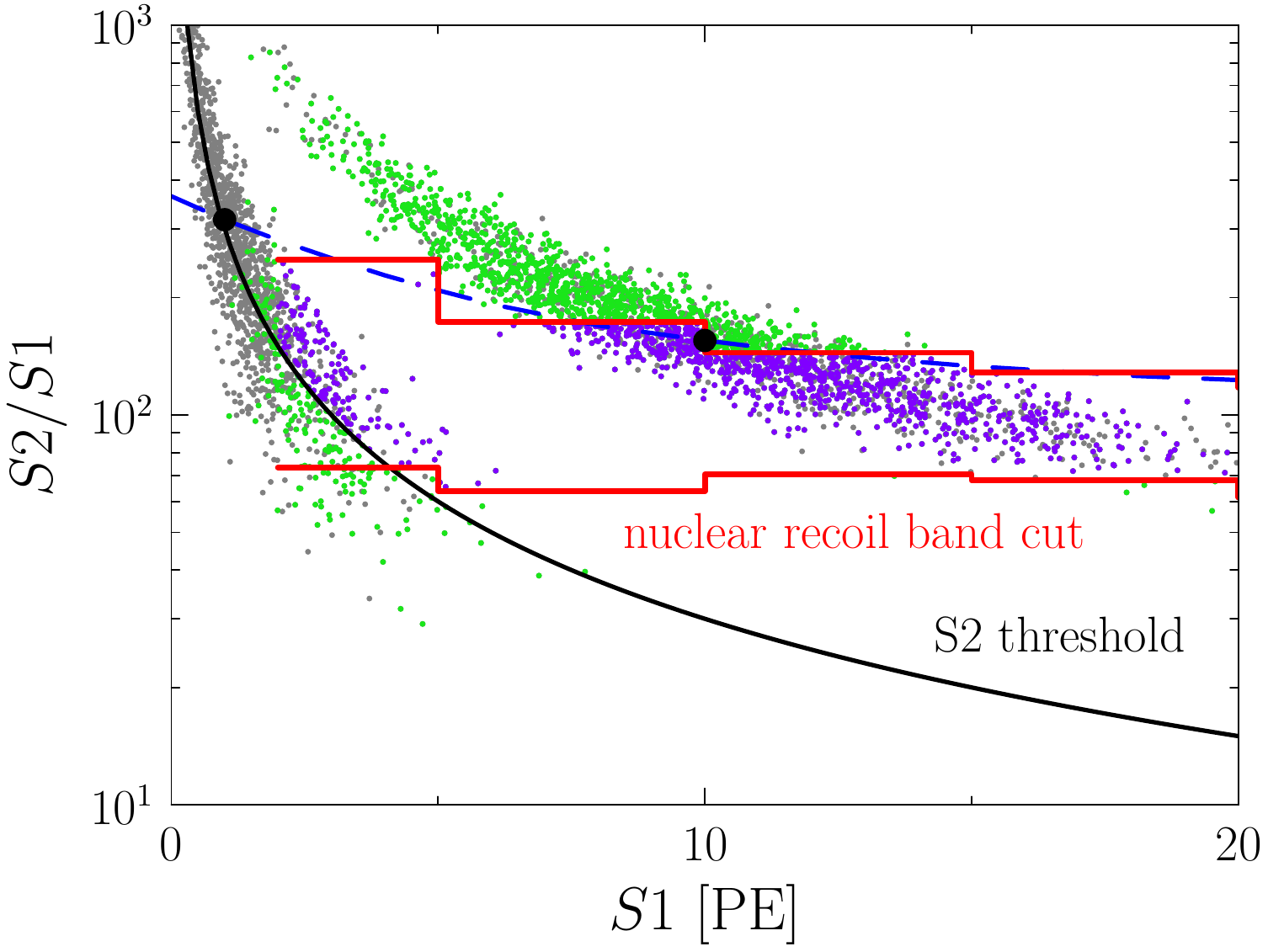}
  \hspace{\stretch{1}}
  \includegraphics[keepaspectratio,width=0.53\textwidth,
                   height=0.35\textheight]{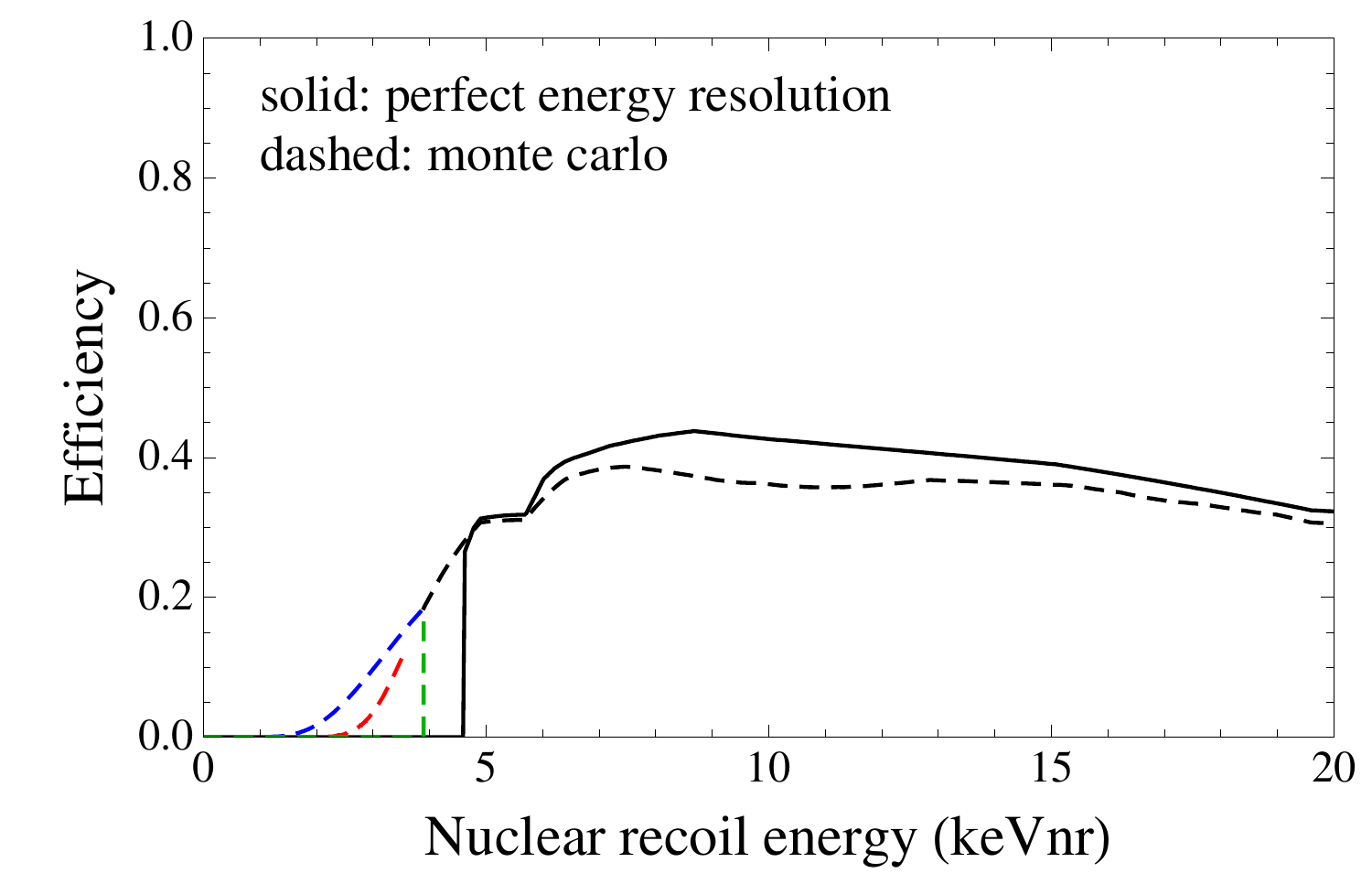}
  \caption[Monte carlo events and efficiency]{
    (\textit{left})
    Monte carloed events in the XENON10 detector for nuclear recoils
    with energies that correspond to average \Sone\ signals (\Soneave)
    of 1.0 and 10.0~PE (black dots).
    Events that fail to be identified via the \Sone\ peak finding
    process are shown in gray.
    Events that are identified but fall either below the \Stwo\ 
    threshold (black curve) or outside the nuclear recoil band cut
    (red curves) are shown in green.
    Events that are identified and pass all cuts (signal events) are
    shown in purple.
    (\textit{right})
    The fraction of events at each recoil energy that will be
    identified and pass all cuts in XENON10 ($2 \le \Sone \le 75$~PE).
    The three colored curves at low energies correspond to the three
    low-energy \Leff\ extrapolations shown in \reffig{LeffModels}.
    }
  \label{fig:MonteCarlo}
\end{figure*}

The standard analysis of XENON10 and XENON100 results uses the two
scintillation signals \Sone\ and \Stwo.  Due to various physical
processes and detector limitations, the observed signals exhibit random
fluctuations about the average expected values at a given nuclear
recoil energy.  The most significant (but not only significant)
contribution to the fluctuations is the Poisson fluctuations in the
small number of PE's produced by the prompt scintillation (only
$\sim$10\% of the prompt photons yield a PE), which leads to a poor
energy resolution at low recoil energies.  We show in the left panel of
\reffig{MonteCarlo} a monte carlo example of the scattering of the
observed \Sone\ and \Stwo\ scintillation signals in XENON10 for recoil
energies yielding an average \Sone\ of 1.0 and 10.0~PE.  We also show
the two main cuts---an \Stwo\ threshold and the nuclear recoil band
cut---and take into account the two-fold PMT detection efficiency
(also known as the \Sone\ peak finding efficiency factor \etaSone).
We use the low-threshold ($\Sone \ge 2.0$~PE) XENON10 results of
Ref.~\cite{Angle:2009xb} in our analysis (see also
Ref.~\cite{Aprile:2010bt} for a detailed description of the XENON10
detector).  We stress that this is a threshold in \Sone, \textit{not}
in \Soneave: as shown in \reffig{MonteCarlo}, a significant number of
events produced at a recoil energy yielding $\Soneave = 1.0$~PE will
have an observed $\Sone \ge 2.0$~PE and pass all cuts, so low-energy
recoils can still contribute to the signal within the XENON10 analysis
region.  To determine the contribution of these low-energy recoils in
XENON10, we monte carlo a large number of events at each recoil energy
and determine the fraction of events that will be both observed and pass
all cuts \cite{Savage:2011}; the results are shown in the right panel
of \reffig{MonteCarlo}.  See Ref.~\cite{Sorensen:2010hq} for further
discussion regarding the fluctuations of the signals in XENON10 and
issues in determining the detection efficiencies.

XENON100 has a lower background than XENON10, but has a higher
threshold of $\Sone \ge 4$~PE \cite{Aprile:2010um,Schumann:2010idm}.
As the full details of the various data cuts in XENON100 are not yet
available, we use for this experiment a simpler model of the efficiency
and impose a $\Soneave \ge 1.0$~PE cutoff as described in
Ref.~\cite{Savage:2010tg}.

\section{\label{sec:Results} Results and Discussion}

\begin{figure*}
  \begin{center}
  \insertwidefig{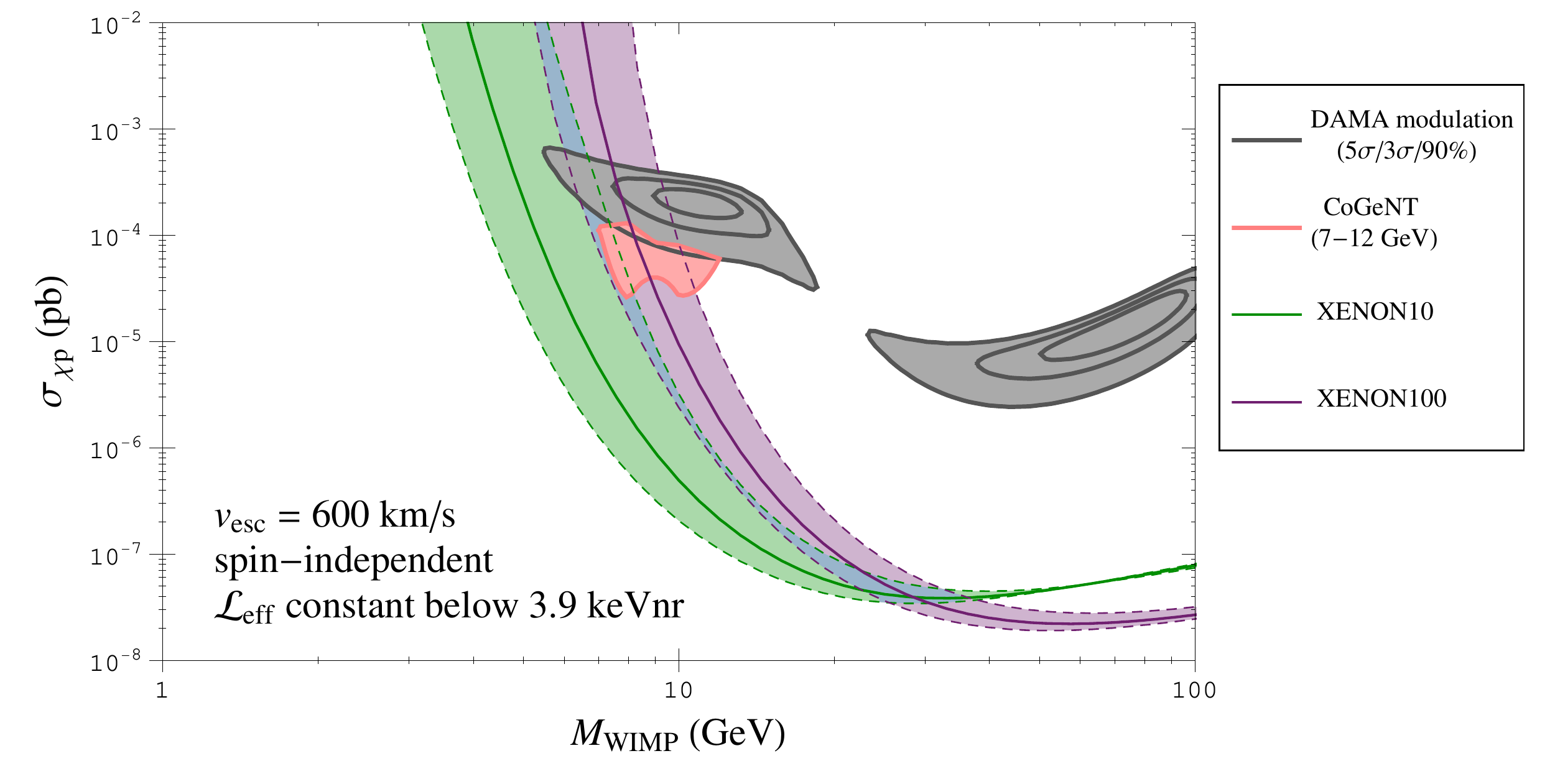}
  \end{center}
  \caption[XENON constraints (flat \Leff)]{
    XENON10 (green) and XENON100 (purple) 90\% C.L.\ exclusion
    constraints for a constant \Leff\ at recoil energies below 3.9~keVnr.
    The solid curves are the constraints using the central values
    of \Leff; dashed curves and lighter filled regions indicate how
    these 90\% constraints vary with the 1$\sigma$ uncertainties in
    \Leff.
    Overlapping XENON10 and XENON100 regions are shown in blue.
    Also shown are the DAMA modulation
    90\%/3$\sigma$/5$\sigma$-compatible regions (gray contours/region)
    and the CoGeNT 90\%-compatible region (pink contour/region
    over 7-12~GeV).
    }
  \label{fig:ConstantLeff}
\end{figure*}

\begin{figure*}
  \begin{center}
  \insertwidefig{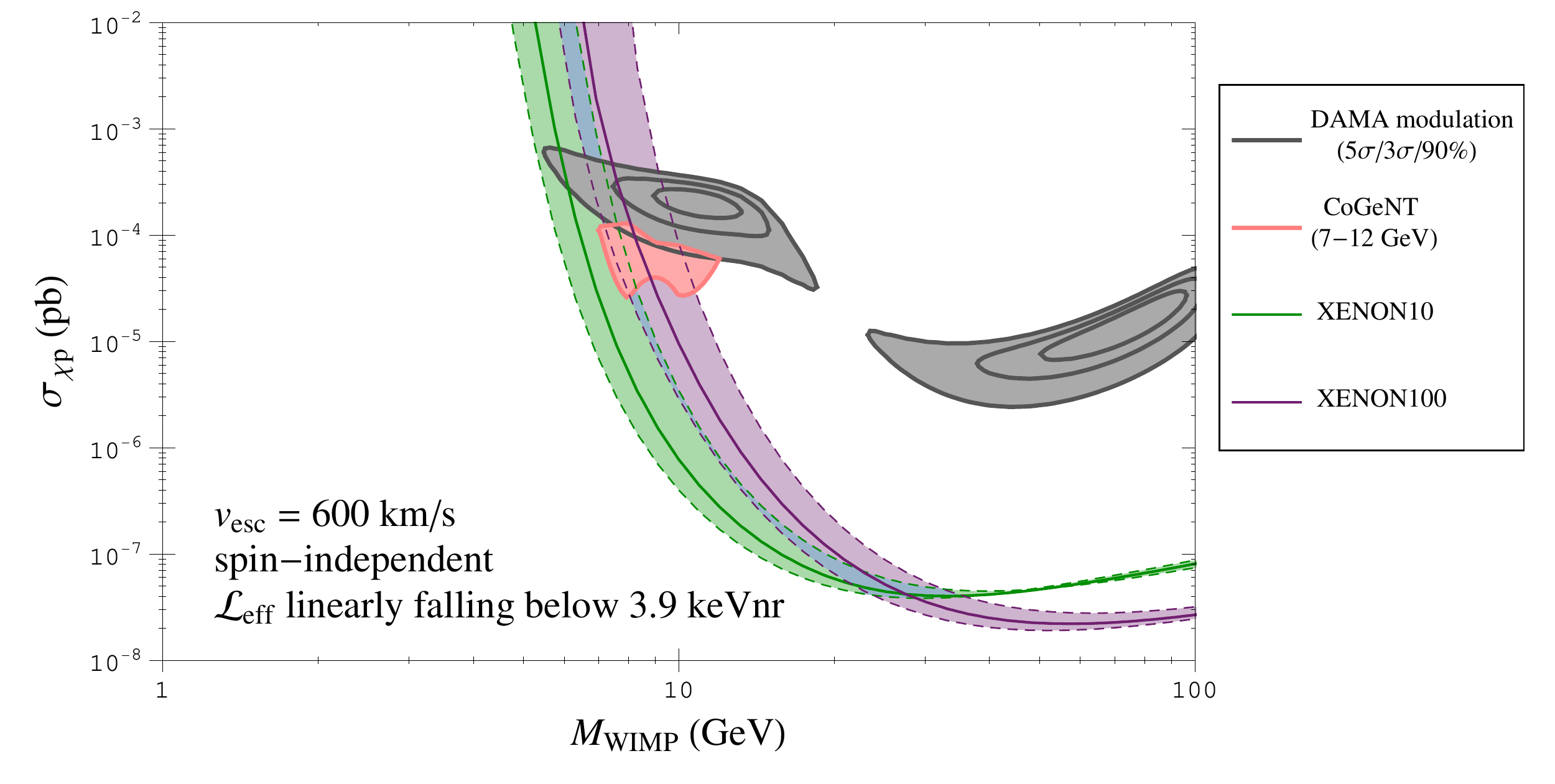}
  \end{center}
  \caption[XENON constraints (falling \Leff)]{
    Same as \reffig{ConstantLeff}, but taking \Leff\ to fall linearly
    to zero for recoil energies below 3.9~keVnr.
    }
  \label{fig:FallingLeff}
\end{figure*}

\begin{figure*}
  \begin{center}
  \insertwidefig{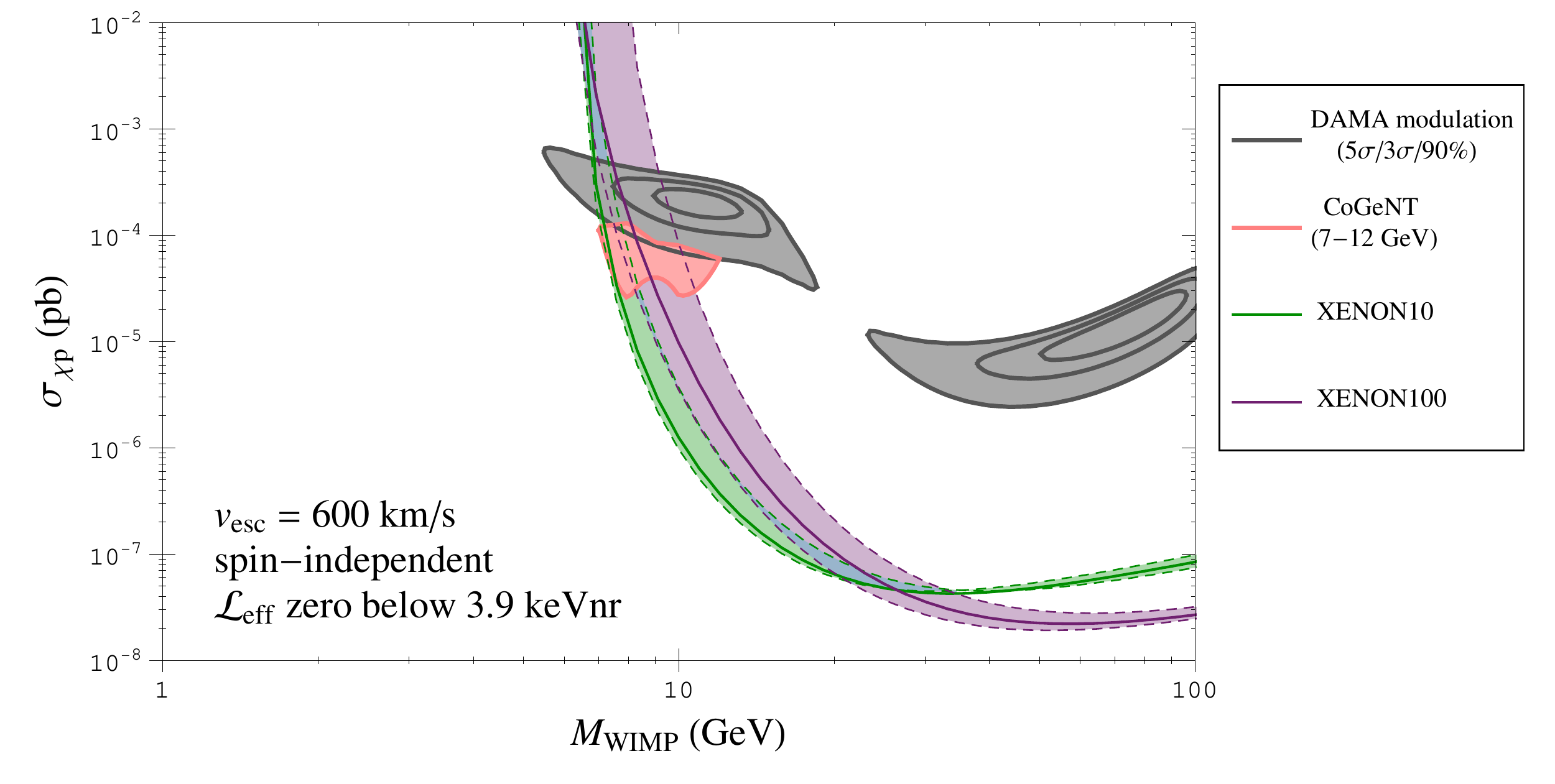}
  \end{center}
  \caption[XENON constraints (zero \Leff)]{
    Same as \reffig{ConstantLeff}, but taking \Leff\ to be zero for
    recoil energies below 3.9~keVnr.
    }
  \label{fig:ZeroLeff}
\end{figure*}

We now examine how the XENON10/100 WIMP mass \& cross-section
constraints are affected by the \Leff\ model.  We consider only
spin-independent elastic scattering and assume an isothermal halo model
as described in Ref.~\cite{Savage:2010tg}.  The results discussed below
apply only to this case; spin-dependent couplings and alternate halo
models may affect the compatibility of various experimental results.

Our main results are shown in
Figs.~\ref{fig:ConstantLeff}-\ref{fig:ZeroLeff},
corresponding to the three cases for the behavior of \Leff\ at low
recoil energies.  The 90\% confidence level (CL) exclusion limits for
the fiducial (central) \Leff\ models are shown as solid curves for
XENON10 (green) and XENON100 (purple); lighter filled regions
correspond to how these 90\% CL exclusion curves vary with the
1$\sigma$ uncertainty bands in the \Leff\ models.
For comparison, we also show the WIMP parameters compatible with
the DAMA modulation \cite{Bernabei:2010mq,Savage:2010tg}
(gray contours/regions, corresponding to compatibility within the
5$\sigma$, 3$\sigma$, and 90\% CLs) and
the region suggested by CoGeNT \cite{Aalseth:2010vx} (pink
contour/region, corresponding to compatibility within the 90\% CL).

The XENON100 constraints are nearly identical for all three \Leff\ 
models.  For the fiducial cases, XENON100 excludes all of the DAMA
3$\sigma$ region, indicating signficant incompatibility between these
two experimental results for this dark matter model.
XENON100 excludes only the 9-12~GeV WIMP mass part of the CoGeNT region,
allowing for 7-9~GeV WIMPs.
When the 1$\sigma$ variations in the Manzur data are considered,
however, XENON100 can exclude as little as only the DAMA 90\% CL region
and almost none of the CoGeNT 90\% region to as much as nearly all of
the DAMA 5$\sigma$ CL and CoGeNT 90\% CL regions.
The similarity of the XENON100 constraints for the different \Leff\ 
models is due to the 4~PE threshold: at recoil energies below 3.9~keV,
where the models differ, $\sim$1 or less PE are expected on average,
so recoils at these energies are unimportant.

Due to the lower threshold of 2~PE, the XENON10 constraints are more
dependent on the \Leff\ model.  The XENON10 constraints for the three
fiducial \Leff\ cases vary significantly, with the low WIMP mass
cut-off varying between $\sim$4~GeV and $\sim$7~GeV. In all three cases,
though, the XENON10 constraints exclude the entire CoGeNT 90\% CL and
DAMA 3$\sigma$ CL regions.  Though the average \Sone\ signal is below
threshold for recoil energies below 3.9~keV (where the \Leff\ models
differ), the random fluctuations in the \Sone\ and \Stwo\ signals
lead to some events at lower recoil energies being observed above the
threshold.  For the light WIMPs in the region of interest for DAMA and
CoGeNT, the number of low-energy recoils is very large and, even if the
fluctuations are small, a significant number of events should still be
observed above threshold.  If the 1$\sigma$ bands on \Leff\ are taken
into account, XENON10 still excludes all of the DAMA 3$\sigma$ region
and all but a narrow band of the CoGeNT region over WIMP masses of
7-9~GeV.

While the ability of the XENON dual-signal (\Sone\ and \Stwo) analyses
to constrain light WIMPs is particularly sensitive to systematic issues
such as the behavior of \Leff, we note that this type of analysis is
optimized more for heavier WIMPs.  An analysis based on only the \Stwo\ 
signal, such as presented by P.\ Sorensen in this conference
\cite{Sorensen:2010hv}, provides much greater sensitivity to light
WIMPs.
However, we have shown here that the standard dual-signal analysis can
still provide stringent constraints on low mass WIMPs even for
conservative models of \Leff.


\acknowledgments
  C.S.\ is grateful for financial support from the Swedish Research
  Council (VR) through the Oskar Klein Centre
  and thanks his collaborators Katherine Freese, Graciela Gelmini,
  and Paolo Gondolo.
  We also thank A.\ Manalaysay, G.\ Plante, and P.\ Sorensen for useful
  discussions.




\end{document}